# Complex optical signatures from quantum dot nanostructures and behavior in inverted pyramidal recesses

G. Juska, V. Dimastrodonato, L. O. Mereni, T. H. Chung, A. Gocalinska and E. Pelucchi

*Tyndall National Institute, University College Cork, Lee Maltings, Cork, Ireland*

B. Van Hattem, M. Ediger and P. Corfdir[*]

*Cavendish Laboratory, University of Cambridge, J. J. Thomson Avenue, CB3 0HE Cambridge, United Kingdom*

**Abstract**

A study of previously overlooked structural and optical properties of InGaAs heterostructures grown on (111)B oriented GaAs substrates patterned with inverted 7.5 µm pitch pyramidal recesses is presented. First, the composition of the confinement barrier material (GaAs in this work) and its growth temperature are shown as some of the key parameters that determine the main quantum dot properties, including nontrivial emission energy dependence, excitonic pattern and unusual photoluminescence energetic ordering of the InGaAs ensemble nanostructures. Secondly, the formation of a formerly unidentified type of InGaAs nanostructures – three corner quantum dots – is demonstrated in our structures next to the well-known ones (a quantum dot and three lateral quantum wires and quantum wells). The findings show the complexity of the pyramidal quantum dot system which strongly depends on the sample design and which should be considered when selecting highly symmetric (central) quantum dots in newly designed experimental projects.

[*]Present address: Paul Drude Institut für Festkörperelektronik, Hausvogteilplatz 5-7, 10117 Berlin, Germany

## I. INTRODUCTION

The practical realization of quantum information processing is still in its infancy. A number of possible implementations and different routes have been studied by the scientific community[1]. Despite that, there is little doubt that the final, practicably usable result will be a complex, integrated large-scale system of networks of quantum bit (qubit) sources, memories, gates, etc. Quantum dots (QDs) are prominent candidates for possible qubit sources or implementations, and their study progresses rapidly[2,3,4,5,6,7]. In this regard, some of us recently demonstrated that certain previous disadvantages of QDs, specifically the low confinement symmetry together with the lack of deterministic positioning could be overcome. This has been made possible thanks to site-controlled InGaAs/GaAs Pyramidal QDs[8] which are grown by Metalorganic Vapor Phase Epitaxy (MOVPE) in inverted pyramidal recesses etched in (111)B oriented GaAs substrate[9] and has allowed, for example, entangled photon emission from a broad *array* of single QDs[10].

Nevertheless, despite the relevance of the system and some years of intensive studies[11,12], we still face an incomplete structural/optical properties characterization and understanding, especially when the InGaAs Pyramidal QD system confined by GaAs[8] instead of AlGaAs barriers is taken into account. Unlike the most widely spread self-assembled quantum-dot systems (e.g., Stranski–Krastanow and droplet epitaxy), the morphology of the heterostructures formed during the epitaxial fabrication of Pyramidal QD samples is more complicated because of the formation scenario of interconnected nanostructures[11]. While recent theoretical results[13,14] describe accurately the interplay between the facet-dependent kinetics of precursors and adatoms, surface chemical properties and geometrical constraints on the central QD formation – clarifying *inter alia* the importance of the self-limited profile of the barriers for the development of the QD geometry[15] – they still lack the capability of modeling the full complexity of the growth process, which involves an intricate facet evolution and nanowire formation mechanism developing over length scales of several microns.

The most widely studied system is the (In)GaAs (QD layer)/AlGaAs (barriers) heterostructure. According to the established picture, an ensemble of interconnecting nanostructures develops[16]. If a nominally thin (a few nanometers) pseudomorphic

InGaAs layer is confined by (Al)GaAs, a quantum dot (QD) is formed at the bottom/centre of the recess. Moreover, the dot is physically connected with three lateral quantum wires (LQWRs) located at the intersections of the adjacent exposed (111)A facets and three lateral quantum wells (LQWs) parallel to the pyramidal recess walls [Figure 1(d)-inset]. A more complicated scenario arises if other than a binary alloy is used as barrier material, including the formation of vertical quantum wire (VQWR) and three vertical quantum wells (VQW), for reasons similar to what has been previously observed in the case of V-groove quantum wires: e.g. the fast diffusion of Ga adatoms in AlGaAs layers tends to enrich the centre of concave corners and induce alloy segregation[14,15].

As will be discussed below, a typical photoluminescence (PL) spectrum of the whole InGaAs/AlGaAs ensemble of nanostructures extends from 1.3 eV to 2.05 eV. According to this seemingly well-established picture each nanostructure has a unique signature of optical properties in the spectrum and they appear ordered on the energy scale[11,12]. Experimentally the QD has been observed systematically at the lowest energy, then typically the emission energy increases when going from the VQWR to the LQWRs, LQWs, VQWs and finally bulk AlGaAs. In this context, the high crystalline quality is confirmed by the atom-like features of the QDs (e.g., narrow linewidth and single photon emission). These allow not only for an unambiguous identification of a QD among the broad spectrum of all the other heterostructures, but also the study of important properties, such as exchange and Coulomb interactions of carriers, confinement potential symmetries and their elevation etc.[17]

In this work, we show that the previously accepted picture of the (In)GaAs/AlGaAs system[11,12] can be altered dramatically as soon as the confinement barrier material is changed to GaAs. This leads to very different optical properties of the QDs and their associated nanostructures, which if properly identified and understood, can allow exploiting the relevant features associated with the pyramidal QD system. We demonstrate that not only the optical properties of the InGaAs/GaAs system differ from that of the InGaAs/AlGaAs one, but also that identification of a QD might not be trivial due to several peculiarities. This complexity arises as (1) the main QD is not necessarily the least energetic feature in the whole spectrum as assumed until now, and (2) its emission can be confused with previously unidentified but usually present (at least in our pyramidal QD system) structures with QD-like photoluminescence. We refer to these structures as corner quantum dots (CQDs), and we demonstrate that they are a building block of the interconnecting nanostructure ensemble and not random features (e.g., due to random alloy disorder or segregation). The presence of three CQDs in each pyramidal recess is indicated by detailed micro-photoluminescence and magneto-photoluminescence experiments. The layout of the results representation in this paper is split into two parts: (1) the first one concentrating on the growth temperature dependent studies, (2) while the second one on the corner quantum dots.

## II. EXPERIMENTAL

The studied $In_{0.25}Ga_{0.75}As$ QDs were grown within inverted 7.5 µm pitch pyramidal recesses etched in (111)B GaAs substrates[8] (a representative SEM image of the recesses after patterning is shown in Fig. 1(a): note that the recesses dimensions are close to the pattern pitch, as it is custom in these cases) . A first batch of six samples (each containing thousands of 0.5 nm nominal thickness QDs) was used to study the evolution of the PL with the alloy concentration. The actual thickness of the QDs depends on the growth conditions – as a rule of thumb, a factor 5-6 thicker dots are to be generally expected. Samples were grown at a nominal thermocouple temperature of ~730˚C, and as it is custom in large pitch pyramidal structures, the pyramidal recess is not fully filled during growth. The indium concentration was varied from 0.15 to 0.65, while the other growth parameters were kept the same. The second batch allowed analysis of the growth temperature-dependence of the PL features: for this, four samples were grown at 640˚C, 670˚C, 700˚C and 730˚C (thermocouple reading, $T_c$; note that this temperature was kept constant for the core layers of the barrier/QD, the actual sample temperature can be estimated as ~600˚C, ~620˚C, ~640˚C and ~665˚C, respectively) using patterned pieces from the same original substrate to avoid any unwanted effects that could possibly appear due to long term growth rate deviations or slightly different substrate processing conditions. The earlier grown sample, used for the CQDs study, had indeed QD photoluminescence features red-shifted by 10-15 meV, compared to its expected replica (0.5 nm nominal thickness, $In_{0.25}Ga_{0.75}As$ alloy composition and 730°C of growth temperature ($T_c$)).

Due to the known (geometrically induced) poor light extraction from the QDs in the as-grown geometry, the substrate was selectively etched away from the samples meant for the growth temperature dependence study[8]. The substrate removal procedure is possible because the samples contain an $Al_{0.75}Ga_{0.25}As$ layer acting as an etch stop layer, inserted at the beginning of the growth and before the actual QD was grown (see Figure 1(b)). The samples were, as it is custom, "surface etched" after growth, i.e. exposed to an isotropic sulfuric acid and peroxide etching for a few tens of seconds to remove unintentional nanostructures formed on the non-growth planar (111)B surface (after a protective photoresist layer was spin coated and exposed to an oxygen plasma to leave only the pyramidal holes covered with the protective layer, leaving the planar (111)B surface free to be exposed to the acid). After the surface etching the photoresist was removed with the appropriate solvent, a few hundred of nm of gold were deposited on top of the MOVPE grown surface. In contrast to previous substrate removal procedures, we used thermocompression gold bonding (instead of black wax or similar as in the past) to attach the supporting substrate to a flat, epiready, GaAs substrate also previously evaporated with gold. After the bonding, which gives mechanical support to the samples, the (111)A back surface of the original (111)B substrate, is immersed for several hours into a selective solution to GaAs, ($NH_4OH:H_2O_2=1:30$), till the tips of the first pyramids appear, and the procedure is stopped. We stress that this procedure creates no significant modification of the optical properties (e.g., neither emission energy shifts nor excitonic pattern changes were observed), while the QD light extraction was increased by several orders of magnitude.

The optical characterization was carried out in a conventional micro-photoluminescence set-up. The samples were cooled down to 7 K in a He closed-cycle cryostat. The excitation source was a laser diode emitting at 635 nm and operating either in pulsed mode or in continuous-wave. The laser light was focused to 1-2 μm$^2$ size spot by a 50x objective with a numerical aperture of 0.55, which enabled access to individual QDs. Linear polarization components were probed by placing a linear polarizer in front of a spectrometer and a motorized half-wave plate directly above the objective to avoid polarization distortion effects in other optical elements.

Photon correlations were analyzed in a time-correlated single photon counting (TCSPC) set-up equipped with 4 silicon avalanche photodiodes connected to a photon counting module. Polarization-resolved cross-correlation curves were built by analyzing single photon detection events in a standard Hanbury Brown-Twiss arrangement[18], by filtering individual excitonic transitions for detection with two monochromators and placing appropriate combinations of quarter-/half-waveplates and polarizing beamsplitters.

The magneto-optical properties of CQDs were investigated using a magneto-optical confocal set-up, which enabled probing photoluminescence at arbitrary angles between the sample growth direction and the magnetic field[19]. The sample was placed in a He bath cryostat with temperature set to 4 K and was excited using continuous-wave He-Ne (632.8 nm) or Ti:Sapphire (730 nm) lasers. The sample was mounted on four piezoelectric stages that allowed translation in three orthogonal directions (x, y, z) and rotation around z axis, where the z-axis is parallel to the optical axis of the set-up. The piezoelectric stages and a short working distance (1.55mm, 0.68 NA) objective coupled to a single-mode fiber were both mounted on a rotatable stage, in order to tilt the sample with respect to the magnetic field. The angle θ between the magnetic field and the z-axis can be varied between -10° and 100°. The optical response of a single dot could be studied in this range, which greatly expands the functionality of conventional magneto-photoluminescence setups.

## III. RESULTS AND DISCUSSION
### III A. QD thickness and associated nanostructures unexpected growth temperature dependence

As discussed above, AlGaAs barriers were mainly used in Pyramidal QD structures till recently. The benefits of using AlGaAs include high confinement barrier and the presence of a VQWR, which acts as an efficient channel for relaxation of charge carriers towards the QD layer, increasing photoluminescence efficiency in the cases of nonresonant optical excitation or electrical injection[12]. On the other hand, using GaAs as a confining barrier instead of AlGaAs has some obvious consequences. First of all, the emission energies of the InGaAs layer nanostructures are reduced due to lower confinement barrier (e.g., the PL of the LQWRs with the same nominal alloy composition and thickness decreases by ~0.2 eV). Second, as alloy segregation effects are not present, the confining barrier is composed of a uniform material (GaAs), reducing confinement disorder. This implies that the QD is no longer in direct physical contact with a VQWR and three VQWs (this impacts obviously the wavefunctions of the charge carriers confined to the QD). Instead, the QD remains connected to only two other types of nanostructures of the same "nominal" alloy composition as the dot itself [20]: three LQWRs and three LQWs [Figure 1(d) inset]. Naturally, the photoluminescence energy of these InGaAs structures is expected to be smaller than the GaAs emission and at values determined by confinement effects (real thickness of the structures and real alloy composition, which we expect generally to be different from the nominal one due to segregation effects). Here, we will concentrate on these InGaAs-related photoluminescence features.

The influence of alloy composition is presented in Figure 1 where representative PL spectra of nominal $In_xGa_{1-x}As$ ($0.15 \leq x \leq 0.65$) nanostructures of 0.5 nm thickness ($T_c=730°C$) are shown. The bandgap shrinkage due to increasing indium concentration is directly reflected in the relative emission energy reduction of all InGaAs nanostructures. Indeed, the QD emission is the most red-shifted (highlighted by a shaded area in Figure 1(d)).

For indium content up to 55%, we observe a typical QD excitonic pattern, which acts as a unique signature for this type of nanostructures. A characteristic feature of these QDs is an antibinding biexciton (XX) transition, i.e. lying at higher energy than the exciton (X) recombination. For higher indium content, strain in the pseudomorphically grown layers is expected to relax through the generation of defects. In this case, we still observe a QD-like emission from $In_{0.65}Ga_{0.35}As$ nanostructures. However, the excitonic pattern shows a random behavior from one QD to another, and we are not reporting further on it here.

The excitonic pattern can be a very sensitive fingerprint for a specific QD structure (QD size, shape, aspect ratio, alloy composition, etc.)[21]. As already mentioned, an antibinding XX is a characteristic feature of all the investigated QDs. This result is consistent with a similar pyramidal InGaAs/GaAs QD system reported in Ref. 22. However, at variance with Ref. 22, we have not observed any XX with a positive binding energy. Here, the QD confinement barrier plays clearly a significant role, as the pyramidal InGaAs QDs within AlGaAs barriers always had a binding biexciton[23]. While the behavior of XX is usually explained in terms of Coulomb interactions between carriers[24], it appears here that the structural details play a major role. For example, we have observed a heavily non-monotonic XX binding energy dependence in a wide range of nominal $In_{0.25}Ga_{0.75}As$ QD thickness (0.45 nm – 1.2 nm) with reduced X and XX energy separation for the largest and the smallest QDs. Incidentally, we believe that such data demands for an advanced theoretical analysis including all the peculiarities of our pyramidal QDs. At the moment we can only confirm an antibinding biexciton as a characteristic feature of our InGaAs site controlled QDs.

While the QD emission could be clearly identified by a number of methods, such as photon correlation and excitation power dependence, the photoluminescence at higher energy could not be solely attributed to LQWRs and LQWs. In fact, even distinguishing between LQWRs and LQWs appears to be more complicated than in structures with AlGaAs barriers where distinct groups of peaks corresponding to the mentioned structures are clearly visible. In thinner structures with lower indium content within GaAs barriers, LQWs and LQWRs are strongly confined and their emission is pushed very close to GaAs and overlap with each other. The distinction becomes clearer in the PL spectrum of thicker and/or higher indium content structures (like in the spectrum of $In_{0.65}Ga_{0.35}As$ in Figure 1 (d)). As discussed below, we assume that the dominant PL mainly comes from the wires.

The latter assumption is based on experimental observations where we studied the impact of growth temperature on the structural and optical properties of pyramidal heterostructures. Growth temperature is one of the key parameters that governs the kinetics of precursors and adatoms that participate in the self-limiting growth in pyramidal (and V-grooved) recesses[25]. The complex evolution of self-limiting profiles is well understood and reliably described within a recently proposed theoretical framework[14]. In short, the geometrical profile of the barrier underlying the QD layer determines the width of the QD. This profile depends on both the growth temperature and the alloy composition. It is thus one of the key parameters that controls the QD properties. The geometrical properties of the QDs are directly reflected in the electronic level structure, which controls the spectrum of the excitonic emission. The latter can therefore to some extent be regarded as a fingerprint of a particular geometry. This offers the prospect of deterministic control of the QD geometry, which is of great practical interest.

Figure 2 summarizes the results of studies on growth temperature dependence. The same 0.5 nm nominal thickness $In_{0.25}Ga_{0.75}As$ within GaAs barriers structure was grown under four different sets of conditions as described in the experimental section: $T_c$= 640˚C, 670˚C, 700˚C and 730˚C. We clearly observe a decrease in the QD emission energy when the growth temperature is increased: for instance, the QD PL redshifts by ~40 meV when the growth temperature is increased from 640˚C to 730˚C. The LQWR emission follows an opposite trend. Surprisingly, the QD and the LQWR emissions overlap in structures grown at 670˚C, and for structures grown at 640°C, the LQWR PL is the lowest energy feature. This observation not only reveals one of the complex features of the system, but raises also questions concerning the element of QD confinement. If the dot is in a direct physical contact with uniform LQWRs at its corners, it is possible that the electronic states in the extended potential are sufficiently finely-spaced in energy to suppress the quasi-zero dimensions features of a QD. However, we always observed QD-like emission from all four samples. This was carefully checked by photon correlation measurements. Moreover, the excitonic pattern with an antibinding biexciton was maintained, irrespective of the growth temperature. More importantly, polarization-resolved correlation experiments proved that all the samples contain, relatively easily found, QDs emitting polarization-entangled photons. We see all this as a definitive proof that our attributions are correct.

An example of such continuous-wave excitation measurement is presented in Figure 3, where representative continuous-wave polarization-resolved second-order correlation measurements are shown. As discussed in a number of contributions, this is a typical protocol to reveal entanglement, which resides in the superposition of two-photon

(biexciton and exciton) polarization states[26]. An expected maximally entangled state is expressed as

$$|\psi\rangle = \tfrac{1}{\sqrt{2}}\left(|L_{XX} R_X\rangle + |R_{XX} L_X\rangle\right), \quad (1)$$

where L and R represent left and right circular polarization states of biexciton (XX) and exciton (X). As a single photon pure polarization state can be expressed as a superposition of two other orthogonal pure polarization states, the maximally entangled state can be rewritten in terms of these states. In linear and diagonal polarization bases, the state becomes

$$|\psi\rangle = \tfrac{1}{\sqrt{2}}\left(|H_{XX} H_X\rangle + |V_{XX} V_X\rangle\right) = \tfrac{1}{\sqrt{2}}\left(|D_{XX} D_X\rangle + |A_{XX} A_X\rangle\right), \quad (2)$$

indicating correlation of counter-polarized photons in a circular basis and correlation of co-polarized photons in a linear (H and V represent horizontal and vertical states) and diagonal (D and A represent diagonal and antidiagonal states) bases. Indeed, this is the type of correlations that has been observed between biexciton and exciton photons as shown in Figure 3. Fidelity of the entangled state was calculated at each delay point $\tau$ using the relation

$$F(\tau) = (1 + C_L(\tau) + C_D(\tau) - C_C(\tau))/4, \quad (3)$$

where $C_{basis}(\tau) = \left(g^{(2)}_{xx,x}(\tau) - g^{(2)}_{xx,\bar{x}}(\tau)\right)/\left(g^{(2)}_{xx,x}(\tau) + g^{(2)}_{xx,\bar{x}}(\tau)\right)$ is the degree of correlation, $g^{(2)}(\tau)$ is the second-order correlation function of co-polarized XX, X and orthogonal polarization exciton ($\bar{X}$). The maximum value of calculated fidelity (0.59±0.04) indicates clearly the presence of non-classical correlations (the maximum limit for ideal classical correlations is 0.5). We emphasize that the photon correlation results shown in Figure 3 were obtained on the sample grown at 640°C. It is not only an unambiguous indication of a QD as a structure as such, even though not mostly red-shifted, but it is a confirmation of the presence of a very similar excitonic pattern and a proof of a high symmetry maintained in the range of the studied growth conditions.

There are several points that need to be addressed regarding our previous PL vs growth temperature investigation: the trends for the emission energy shift observed for QDs and LQWRs and the fact that LQWRs are the type of structures emitting at the lowest energy when using reduced growth temperatures. To verify the influence of growth temperature on the QD morphology, an AFM cross-section image of GaAs (darker regions) layers grown at $T_c$ 640˚C, 670˚C and 700˚C was taken [Figure 4]. GaAs layers were thick enough (100 nm) to develop a self-limited-growth at each corresponding temperature. The width of GaAs self-limiting profile (or otherwise the QD width determined by the underlying layer) was measured at the interface of GaAs and $Al_{0.3}Ga_{0.7}As$ markers (as indicated by white dashed lines in Figure 4). The width was clearly increasing from ~30 to ~70 nm, when the temperature was increased from 640 to 730˚C, respectively (note that the latter value was known and already measured in Ref. 27) [Figure 4 (b)]. The evolution of morphology of such patterned surfaces (despite not being reported before) follows the trend predicted using the modeling described in Ref. 14, confirming the accuracy of the modeling results. On the other hand, while the width of the QD decreases by more than a factor of two at the lowest temperature, the influence of growth temperature on a QD alloy itself (segregation, exact QD thickness and shape) is not obvious.

      We can try to speculate on these points using growth analysis theory and experimental results based on the AlGaAs alloy. In general, one expects InGaAs to be analogous in the physics involved. It is well established that in AlGaAs structures an enrichment of gallium is present at the central axis of the pyramidal recess due to the more diffusive nature of gallium as compared to aluminum. Similarly, as indium atoms are more diffusive than gallium, segregation of indium is also expected. Accordingly, this trend was observed in nanowires grown in V-groove shaped patterns[25]. To understand how segregation phenomena evolve with temperature, some modeling of (Al)GaAs alloy segregation has been carried out, based on the phenomenological and well tested theoretical framework presented in Ref. 14. The result of such calculations shows no dependence of the relative gallium concentration along the vertical axis of the pyramid on the growth temperature[28]. If we assume a similar scenario for the InGaAs alloy, we can then exclude, possibly, the segregation effect as the main source of blue-shift with decreasing growth temperature. At this stage we also tend to neglect a significant change in the QD thickness with growth temperature, as the small relative change observed in Figure 4 is coherent with the well-known, expected growth rate increase with dot position along the central recess. Nevertheless, we caution the reader, that in our images the dot layer has been grown thicker for the needs of avoiding resolution issues with AFM and

that the actual QDs are much thinner. As a consequence we cannot exclude *a priori* that over the first few nm the growth process proceeds in a qualitatively different way. By taking into account all three assumptions ((*1*) self-limiting profile shrinking at lower temperatures, (*2*) somehow constant alloy composition and (*3*) almost similar growth rate for the QD), we can attribute the QD energy increase at the lowest temperature growth to the increased confinement associated to the shrinking of the QD size.

We can tentatively discuss the opposite trend observed for the LQWRs relying on similar arguments. Within this scenario (the relative gallium concentration in the middle is constant and the QD volume is reduced – roughly the same thickness, narrower base) and invoking the mass conservation in the vicinity of the dot (the total flux we send and amount of precursor decomposition are the same), a lower growth temperature would result in an enhanced relative amount/concentration of gallium on the lateral facets/LQWRs, not necessarily in a uniform fashion, as the reduced incorporation of group III elements in the QD will be compensated elsewhere. The increased gallium relative concentration could lead to a redshift of the LQWRs emission, even if it is unlikely to impact the full length of the wires (of the order of a micron). However, we need to consider the effect of the self-limiting profile for the LQWRs: the lateral profile along the edges of the template is expected to narrow down as the growth temperature is reduced (pretty much for the same reasons as for the QD) and this effect opposes the increased gallium (segregation) amount. If we assume the latter (gallium increase/segregation) to be dominant on the profile variation, then the overall effect of the lower growth temperature could result in a "reduced" redshift of the LQWRs (note that the magnitude of the energy blueshift seen for the QDs is larger than what is observed for the LQWRs).

Moreover, to explain the observed relative energy position for the LQWRs and QD emissions in nanostructures grown at 640˚C, the profile of a single LQWR has to be considered. It was already demonstrated that in the GaAs/AlGaAs Pyramidal QD system, the shape of GaAs LQWRs is not uniform along their axes, but tapered towards the center of the pyramid[11,29]. InGaAs LQWRs are expected to exhibit an analogous structure. As a consequence, the LQWR PL at low excitation powers is mainly emitted from regions close to the top corners of the pyramidal recess, while a QD remains confined by the thinnest segments of the LQWRs. This would also be consistent with the persistence of a QD confinement, despite the presence of other nanostructures showing lower energy emission.

We stress that these are mere conjectures, which will need to be confirmed by an extensive transmission electron microscopy analysis. Unfortunately, such a task might reveal quite challenging in a three-dimensional system like the one investigated in our manuscript. On the other hand, a more complex scenario involving a different decomposition vs indium diffusion/segregation as a function of temperature in the case of "center" of the pyramid and "V-groove alike" sides (in the former system the competition is between (111)B and (111)A surfaces, while in the latter it involves (100)+(311) vs (111)A surfaces ) might actually be the source of this unexpected behavior. More work is ongoing in our group to test this possibility, and will be published should we obtain an unambiguous answer to the question.

**III B. Corner QDs**

A further complexity of the Pyramidal QD system is revealed by identifying and demonstrating the existence of a previously disregarded type of structure. In Figure 1(d), we have highlighted using open rectangles specific energy ranges of the PL spectra of $In_{0.25}Ga_{0.75}As$ and $In_{0.45}Ga_{0.55}As$ nanostructures. These QD-like spectra appear for all pyramidal QD structures under appropriate optical pumping conditions and are not accidental to a specific measurement or sample. Moreover, we discard the fact that these sharp QD-like peaks (sometimes as narrow as a few tens of µeV) could be related to excited states of the QDs. Using excitation power-dependent measurements, it appears that these emission peaks can be often observed using an excitation power lower than what is required for the main QD emission. These peaks were also observed in some as-grown samples showing no main QD emission. We acknowledge that it is very common in apex-down geometry (or as-grown) InGaAs/GaAs pyramids not to see the main QD photoluminescence at all due to the very poor QD emission extraction. A closer study shows that these peaks have typical QD properties: multiexcitonic transitions are identified by excitation power-dependent and time-resolved measurements, single photon emission is demonstrated and the relevant fine-structure splitting is measured and characterized. All the methods used here suggest that a "new" QD system has been found. To understand more about these QDs and their relation with the previously known nanostructures we present a detailed analysis of their emission properties in the following of the manuscript.

**III C. Corner vs main dots characterization by polarization anisotropy**

A rarely used but efficient way to characterize a QD is to probe its emission from the cleaved side[30,31]. Pyramidal QDs are favorable for such measurements due to site-controlled positioning of the QDs, which allows an easy access to a significant number of QDs even in side-view. This is achieved by cleaving the sample and placing in the cryostat facing the edge up (see the sketch in Figure 6(a)). By this method, we have studied a particular $In_{0.25}Ga_{0.75}As$ sample grown at 730°C that showed systematically the above-mentioned QD-like emission peaks. From now on, we refer to these emission peaks as corner quantum dots (CQDs) due to reasons that will become clearer in the following. The initial characterization of the peaks in top-view showed very clear QD properties: exciton-biexciton dynamics, fine-structure splitting reflected in both transitions. We show in Figure 5 a PL spectrum from a representative CQD together with the emission of the main QD, taken under the same excitation conditions. We could only rarely observe the PL from both the CQD and the main QD using the same excitation power. This is due to the already mentioned poor PL extraction efficiency for the main QD, which forces us to use relatively high pumping conditions. Therefore, the spectrum taken on the main QD does not reflect its usual optical quality, which is typically characterized by a narrow linewidth. In a previous report based on the same sample, the minimum linewidth of the main QD from a bright sample where the substrate was removed was typically found to be smaller than the resolution of the experimental set-up (18 μeV) even though, at that time, the interpretation of CQDs states was missing and some conclusions were inappropriate[8,32].

The emitted photoluminescence was analyzed with a linear polarizer placed after the rotating half-wave plate. As presented in the inset of Figure 5, the PL from both the main QD and the CQD show strong polarization anisotropy. Polarization anisotropy is an expected result as it can be explained in the terms of a radiant dipole placed in a flat QD growth plane. It is a valid picture if ground state transitions between heavy-hole and electron are observed[30,33]. Advanced analysis and theoretical models demonstrated that polarization anisotropy can be a unique tool to probe transitions that involve different types of holes[30,34]. Here, the maximum intensity of the main QD (exciton transition) is obtained with the linear polarizer perpendicular to the growth axis. In contrast, it is nearly suppressed when the polarizer is parallel to the growth axis. The polarization anisotropy of the main QD is characteristic of QD ground-state transitions involving heavy-hole exciton recombination. Coming to the CQD emission lines, they exhibit unusual polarization anisotropy properties that cannot be understood in terms of heavy-(light-) hole properties. For instance, the emission intensity of the representative CQD in Figure 5 is maximum when the polarizer makes an angle of ~35˚ with the growth axis.

The results of a systematic study of the polarization anisotropy properties are shown in Figure 6(c). The maximum intensity values are clearly bunched into three groups. As expected, the main QD PL is always systematically polarized perpendicular to the growth axis. Regarding the CQDs, they show maximum emission intensity when the polarizer makes an angle of about 35˚ or 160˚ with the growth axis. The relevance of these angles becomes clearer, if the geometry of the pyramidal recess is taken into consideration. Figure 6(b) shows the projection of a measured pyramid cleaved in the plane perpendicular to one of the facets (the real orientation during the measurements) and the polarization analysis scheme (the reference point and rotation direction). Considering 0˚ axes being perpendicular to the growth plane, 35˚20', 90˚ and 160˚30' are the angles that match the directions along the three projections. Since all three LQWRs are located as wedges between adjacent pyramidal facets, 35˚20' and 160˚30' also represent projections of the LQWRs. Consequently, the results in Figure 6(c) suggest a strong relation between CQDs and LQWRs. In a way similar to what happens for the central QDs, LQWRs grow on a self-limiting profile that develops between the adjacent pyramid walls. As a result, the shape of LQWRs is usually flat. The polarization anisotropy observed for CQDs suggest that they are flat as well and that their plane matches the flattening of the corresponding LQWR.

We would like to address the correlation observed between the spatial excitation point and the PL intensity of the CQDs. In side-view experiments, the emission intensity from CQDs is maximum when the laser is focused close to the surface of either side of the pyramid projection (see the spots L and R indicated in Figure 6(b)). When the CQD emission is collected from the point R, its polarization axis makes systematically an angle of ~35˚20' with the sample growth axis. Correspondingly, this angle is about 160˚30' when the laser is focused in the vicinity of point L.

These observations are consistent with conventional top-view geometry measurements. As shown in Figure 7, it was relatively easy to find pyramids showing simultaneously several CQD peaks, when the laser is focused at the center of the pyramid. Moving the laser spot to the pyramid corners, the PL intensity of a single pair of CQD peaks (exciton and biexciton-related lines) increases while the emission intensity of the other peaks decreases. Note, however, that all spectra show at least weak signatures of each CQD.

Finally, it is possible to observe the photoluminescence of single CQDs from cleaved fractions of pyramidal recesses that do not contain the central part, where the main QD is located. This confirms the significant spatial separation between the CQDs and the central QD. We stressed that we have taken special care to make sure that the PL from the main QD and the CQDs originate from the same pyramid and not from the adjacent. According to these observations made in this Section, the following points can be derived: there are three additional QD-type structures in our pyramidal recesses, they are closely related to the LQWRs and they are located close to the top of the sample and to the corners of the pyramidal recesses.

**III D. CQD signature by FSS measurements**

In this Section, we characterize the symmetry of the confining potential of CQDs by measuring the fine-structure splitting of their exciton emission (FSS). A FSS is observed for the majority of QDs systems in the form of a splitting of the exciton into two orthogonally linearly polarized states. The FSS results from the electron-hole exchange interaction and its magnitude can be seen as a measure of the symmetry of the QD. As any change in the confining potential symmetry, e.g., caused by physical QD elongation[35,36], strain[37], random alloy segregation[38], affects the magnitude of the FSS, it is an important parameter characterizing the QDs. In Figure 8, we plot the evolution of X and XX peak emission energies as a function of the polarizer angle for a given CQD. The angle-dependences of X and XX emission energies are sinusoidal and show as expected a 90° phase shift one with each other. We deduce for this specific CQD a FSS of ~21 μeV. We also display in Figure 8 FSS measured on more than 50 CQDs. All CQDs exhibit a measurable FSS with a mean value of ~13 μeV. While the mean FSS value of the present CQDs is rather small value regarding other QD systems, it is significant when compared to the FSS exhibited by the central dots (e.g., 3.3±2.1 μeV for QDs which are nominally thicker (0.8 nm)).

One of the main reasons for a non-zero FSS is QD elongation along a particular crystallographic direction[35,36]. The polarization axes for X and XX usually match the QD elongation directions. For instance, from the FSS measurement shown in the inset of Figure 8, the polarization axes of the corresponding CQD make an angle of 73 and 163° (can be referred to as H and V). We display in Figure 9(a) the orientation of the polarization axes for 14 pyramids containing 20 CQDs in total. Since both axes are orthogonal, we only display the angle at which the lower-energy exciton state is observed. The values obtained for the orientation of the polarization axis are bunched into three groups, each of these groups consisting of angles values scattered next to 30°, 90° and 150°. Similarly to side-view measurements, the relevance of these angles can be understood by taking into account the orientation of the pyramids within the measurement set-up. As presented in Figure 9(b), 30°, 90° and 150° correspond to the angles where the polarizer is parallel to the top-view projections of the LQWRs. The top inset of Figure 7 shows how the polarization axes correlate with the CQD location. All three CQDs were observed from this pyramid and their polarization axes were found to be 33°, 84° and 159° as indicated in the insets. It is obvious that the polarization axis of each CQD nearly matches the LQWR axes. This is the case for all measured CQDs. Consequently, our observation confirms that CQDs are located within LQWRs. In addition, the one-dimensional nature of LQWRs is likely to be responsible for the FSS of CQDs. In particular, we believe that it arises from a small elongation of the CQDs along the LQWRs.

## III E. Characterization of CQDs by magneto-photoluminescence

The evolution of the CQD exciton energy in a magnetic field ($B$) can provide additional information on the symmetry of these nanostructures. In the presence of $B$, the exciton PL splits and its energy shifts following a parabolic dependence with $B$. The Zeeman interaction between the spin of the exciton and $B$ causes to the emission peaks a splitting and an energy shift linear with $B$; diamagnetic effects induce an energy blue-shift quadratic with $B$ [39]. The measure of the diamagnetic coefficient ($\gamma$) is of particular interest, as this quantity is proportional to the exciton correlation length in the plane perpendicular to the direction of $B$ [40,41,42]. Measuring $\gamma$ for various orientations of $B$ thus allows extraction of the shape of the investigated QD, a method that has already been applied successfully to interfacial QDs,[43,44] Stranski-Krastanov QDs,[45] and quantum disks in nanowires[46].

We show in Figure 10 the magneto-PL for two distinct CQDs (namely CQD1 and CQD2 located in different pyramidal recesses) and for various values of the angle $\theta$ between the direction of $B$ and the sample growth axis [Figure 10]. We emphasize that for both CQD1 and CQD2, we orientated the sample so that the axis of the LQWR that corresponds to the CQD investigated remains in the plane defined by the sample growth axis and the direction of $B$ [Figure 11]. In addition, while for CQD1 the LQWR axis makes an angle ɸ ≈ 35°20' with the sample growth axis [Figure 11(c)], the angle between the growth axis and the axis of the LQWR of CQD2 is -ɸ [Figure 11(f)]. At zero $B$, the exciton energies of CQD1 and CQD2 are 1455.0 and 1456.8 meV, respectively, and we attribute the emission at 1453.6 and 1455.2 meV, respectively, to the recombination of the biexciton. We deduce that the biexciton binding energy for these two CQDs are 1.4 and 1.6 meV, respectively, in agreement with what reported in Figure 8. When applying $B$ with an angle $\theta = 0°$, we observe that the exciton emission of both CQD1 and CQD2 splits into four lines. We extract for these two CQDs $\gamma$ = 19.2 and 20.1 µeV/T$^2$, respectively. Now, when increasing $\theta$ from 0 to 90°, we find that $\gamma$ shows a non-monotonic $\theta$-dependence and we observe two different behaviors for CQD1 and CQD2. As shown in Figure 11(b), $\gamma$ for CQD1 increases from 19.2 to 28 µeV/T$^2$ when $\theta$ is increased from 0 to 54°, and then decreases down to 24.3 µeV/T$^2$, when $\theta$ is increased up to 90°. In contrast, for CQD2, γ exhibits a decrease followed by an increase when $\theta$ goes from 0 to 90° [Figure 11(e)]. As already indicated, CQDs are presumably flat and oriented within LQWRs so that the plane of each CQD matches the flattening of the corresponding LQWR. Consequently, we expect to measure the smallest (*largest*) γ when *B* is parallel (*perpendicular*) to the LQWR axis. For the pyramid geometry shown in Figure 11(a), the LQWR embedding CQD1 is perpendicular to *B* when $\theta = 54°40'$ (i.e. π/2 - ϕ). In agreement with this picture, we measure the maximum γ-value when $\theta = 54°$ [Figure 11(b)]. Coming to the case of CQD2, the wire is parallel to *B* when $\theta = \phi$ [Figure 11(e)]. We measure accordingly the minimum γ-values for $\theta = 30°$. We plot in Figure 12(a) γ for CQD1 and CQD2 as a function of θ and (θ - 2ϕ), respectively. We observe that the γ-values measured for CQD1 and CQD2 follow a similar trend. Assuming that the electron and hole masses are the same for CQD1 and CQD2, this indicates that from one pyramid to another, the CQDs show similar shape and dimensions.

We investigated more closely the evolution of the CQD emission pattern when *B* is parallel or perpendicular to the LQWR axis. The former and latter experimental geometries correspond to the data taken on CQD2 for $\theta = 30°$ [Figure 10 (b)] and on CQD1 for $\theta = 54°$ [Figure 10(a)] respectively. When *B* is parallel to the LQWR, four emission lines are resolved, as a result of mixing between the dark and the bright exciton states. This indicates that if the CQD exhibits a high-symmetry axis, this is not aligned along the LQWR axis, In other words, the present experimental geometry (*B* parallel to the LQWR axis) can be seen as an analog to the Voigt geometry for Stranski-Krastanov QD. In contrast, when *B* is perpendicular to the CQD1 plane, we resolve only two emission lines [Figure 10(a)]. We assume that in this case, *B* is aligned along a high-symmetry axis of the CQD. Using a similar analogy as previously, this orientation would match the conventional Faraday geometry for Stranski-Krastanov QDs. In this situation, *B* does not hybridize the bright and the dark exciton states and only a doublet is observed (as a notable exception, we mention QDs with a $C_{3v}$ symmetry, where four emission lines are observed when *B* is parallel to the $C_3$ axis [47,48,49]).

We have also extracted the isotropic exchange splitting $\delta_{iso}$ for both CQDs. We show in Figure 10 (a,b) the B-dependence of CQD1 and CQD2 exciton energies for various values of θ. From the splitting at zero B between the bright and dark exciton states, we get the isotropic exchange splitting $\delta_{iso}$. We obtain $\delta_{iso}$ = 125 and 149 μeV for CQD1 and CQD2, respectively. As shown in Ref. 39, $\delta_{iso}$ is inversely proportional to the QD volume. The CQDs studied in our work exhibit accordingly larger γ and smaller $\delta_{iso}$ values than what has been reported for Stranski-Krastanov, pyramidal and droplet epitaxy QDs in Refs. [45,48,49,50]. Finally, we plot in Figure 12(b) the angle-dependence of the exciton Landé factor (g) measured for the bright states of CQD1 and CQD2 as a function of θ and (θ−2ϕ), respectively. We have extracted from the magneto-PL of CQD2 taken at θ = 30° that the g-factor for *B* parallel to the LQWR axis is $|g_{//}|$ = 1. Coming to the g-factor for B perpendicular to the LQWR axis ($g_\perp$), our spectral resolution (150 μeV) is too low to resolve any splitting between the two bright states, leading to the conclusion that $|g_\perp|$ ranges between 0 and 0.5. Still, it is clear from Figure 12(b) that g is anisotropic. The electron Landé factor in III-V semiconductors being almost isotropic,[45,50] we attribute tentatively the anisotropy of |g| to some anisotropy of the hole Landé factor. The latter may result from heavy-hole light-hole mixing[51,52] and also from some spreading of the hole wave function in the barrier of the LQWR in the direction parallel and perpendicular to the LQWR axis[45].

To conclude this section, the values taken by *γ* and $\delta_{iso}$ are consistent with the picture of CQDs being relatively large and flat disks lying in the LQWRs plane. The anisotropy observed for *γ* confirms the asymmetry of CQDs. This structural feature allow probably for a significant heavy-hole light-hole mixing which results in a small $g_\perp$. Remarkably, we have observed similar *γ* and g-factor values for CQDs present in two different pyramids. As the g-factor of the hole and therefore of the exciton, for example in the case of InAs QDs, is extremely sensitive to the geometry of the QD[53,54], our observation leads to the conclusion that the CQDs from one pyramid to another show comparable shape and structural properties.

### IV. A few more speculations on CQDs

The analyzed sample with CQDs presents a very good uniformity and density of these states, observed nearly from half to nearly all of the pyramids, depending on the sample area that is probed. A lower density of CQDs with more random properties (emission energy, excitonic pattern, fine-structure splitting values) has been observed in samples with different alloy concentration and/or thickness (for example, see the spectrum of $In_{0.45}Ga_{0.55}As$ structures in Figure 1(d)). Some of these samples are analyzed in Ref. 32 despite the wrong identification of the type of nanostructures (i.e. they have been considered the central dots).

The fact that in some samples this feature is not pronounced or missing, is an issue that requires discussion. Probably this can be attributed to the first post-growth processing step, which is routinely performed (surface etching) and that is meant to etch away irregular growth structures close to the surface of the pyramidal recesses. The procedure is essential, as the mentioned irregularities lead to a broad photoluminescence background overlapping with the systematic features from the pyramidal recess. The lack of a precise control of the post growth processing procedure might result in "etched away" CQDs, as they are likely to be close to the surface. This would explain why these structures have been overlooked till now.

We need also to clarify as a final remark, that despite the obvious proof of their existence, the structural origin of CQDs is still obscure, and more work is needed to understand their formation: while more systematic work is needed, at this stage we have no evidence to believe corner dots are especially induced by a specific recipe (e.g. specific QD layer thickness or growth temperature). We believe instead that their presence is a generic feature in our structures.

## V. Conclusions and summary

In this work, a detailed study of a few, previously overlooked or unidentified structural properties of InGaAs within GaAs heterostructures grown in large 7.5 µm pitch pyramidal recesses was presented. Among an ensemble of highly symmetric nanostructures (proved by entangled photon detection) the central QD was identified as not being necessarily the least energetic feature, in clear contrast with the previously accepted picture. The emission energy crossover between the central QD and the lateral quantum wires was tentatively attributed to the peculiarities of the epitaxial growth on non-planar substrates at different temperatures and their effect on the dot base and confinement, somehow affecting the dots and lateral wires differently (not excluding as a possibility that of different indium behavior, even if unexpected). Generally speaking, our results point towards the need of further microscopy analysis to clarify the open issues, together with theoretical modelling.

A new type of nanostructure in the ensemble was also identified. The presence of three additional corner quantum dots (CQDs) in the pyramidal nanostructure was supported by in-plane and top-view photoluminescence measurements and by magneto-photoluminescence experiments. We show that these CQDs form in the LQWRs. They present a mean fine structure splitting of 13 µeV that arises presumably from the elongation of the CQD along the LQWR axis.

# Figures

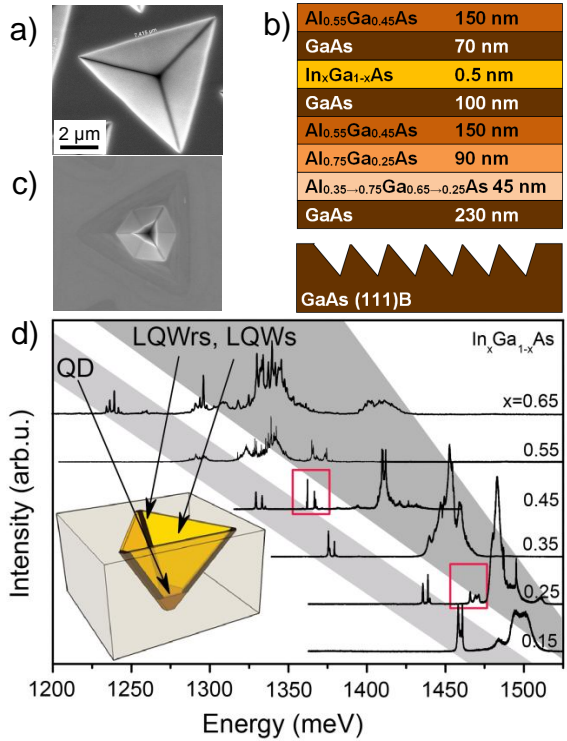

**Figure 1.** (Color online) (a) A representative SEM image of a recess etched in GaAs substrate before the growth procedure. (b) The schematics of a layered structure grown in pyramidal recesses (nominal thickness values). (c) A representative SEM image of a recess after the MOVPE growth. (d) Representative photoluminescence of $In_xGa_{1-x}As$ ($0.15 \leq x \leq 0.65$) 0.5 nm nanostructures grown in 7.5 µm pitch pyramidal recesses at 730°C $T_c$. The sketch of the organization of nanostructures in a recess is shown in the inset. The red squares highlight additional QD-like peaks which we refer to as corner QDs.

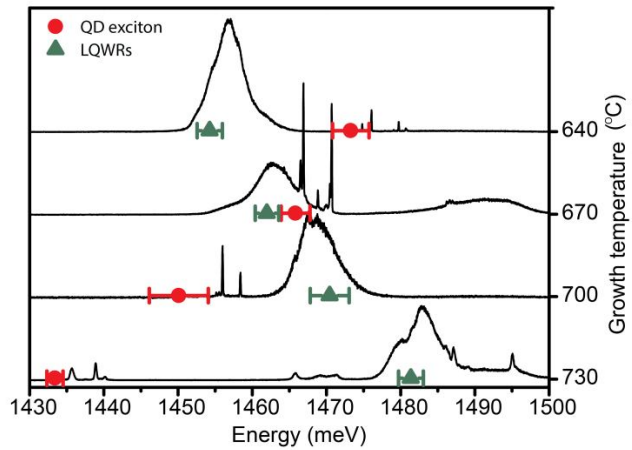

**Figure 2.** (Color online) Representative photoluminescence spectra and average LQWRs and QD neutral exciton emission energy of 0.5 nm $In_{0.25}Ga_{0.75}As$ nanostructures grown at four different temperature values. The error bars are standard deviations obtained from the measured peak distributions.

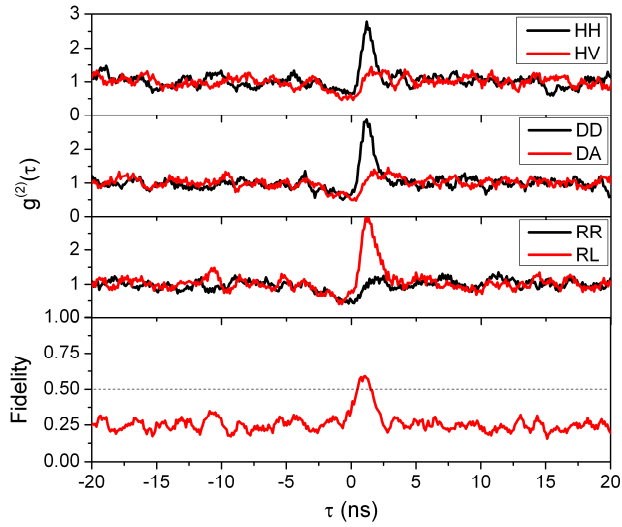

**Figure 3.** (Color online) Continuous wave polarization resolved second-order correlation curves taken in linear, diagonal and circular bases. Fidelity vas calculated using these curves and the maximum value was found to be 0.59±0.04, indicating polarization entanglement (>0.5).

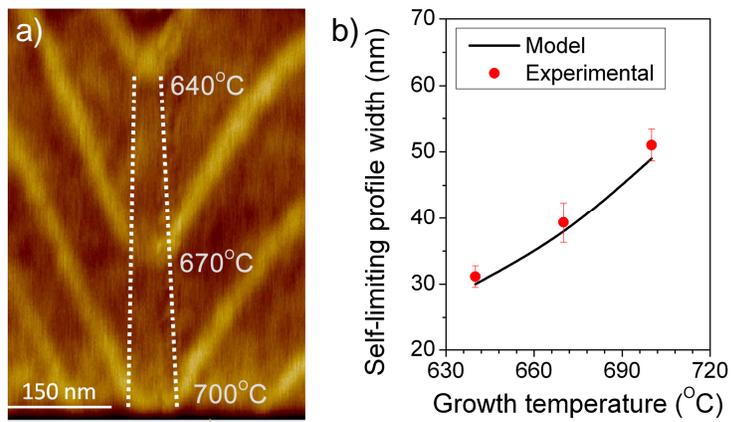

**Figure 4.** (Color online) (a) AFM cross-sectional images of GaAs layers (darker) grown at different temperature values. The evolution of the self-limiting profile is shown by the dashed lines. (b) Self-limiting profile width as a function of the growth temperature (symbols). The solid line shows the result of the fit obtained using the model described in Ref. 14.

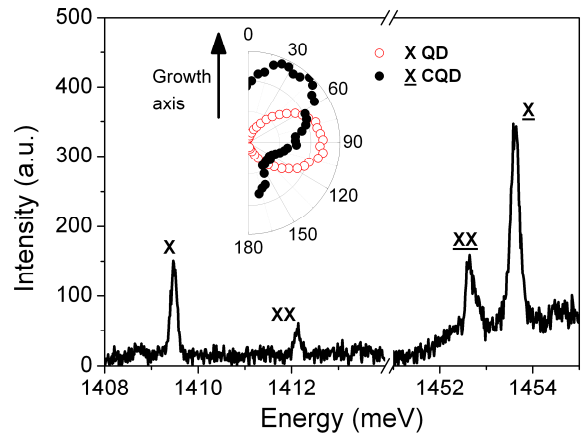

**Figure 5.** (Color online) Representative spectrum taken in side-view where both QD-like features (the main central QD and a single CQD) are visible simultaneously using the same excitation conditions. (Inset) Polarization anisotropy of QD and CQD when probed from the side.

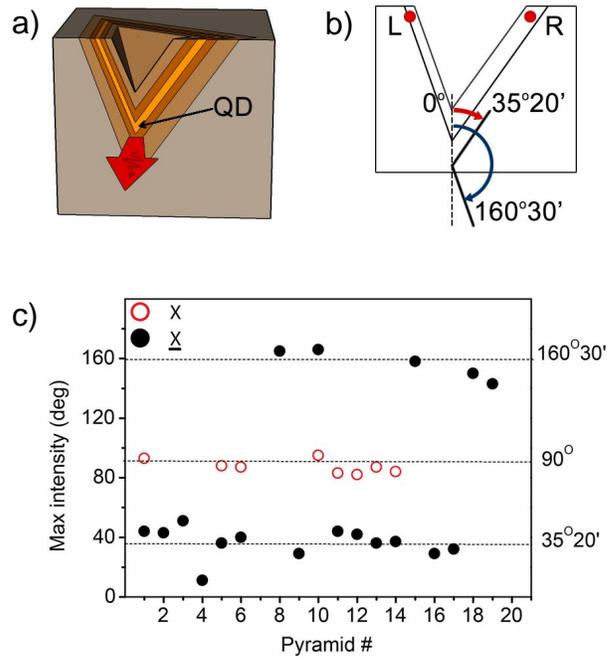

**Figure 6.** (Color online) (a) Sketch of a side-view experiment. PL is collected from the cleaved edge. (b) Projection of a cleaved pyramidal recess as in the experiment. The orientation of the projections of LQWRs in respect of the zero axis is described by the presented angle values. (c) Polarization anisotropy maximum distribution of QDs and CQDs probed from the side.

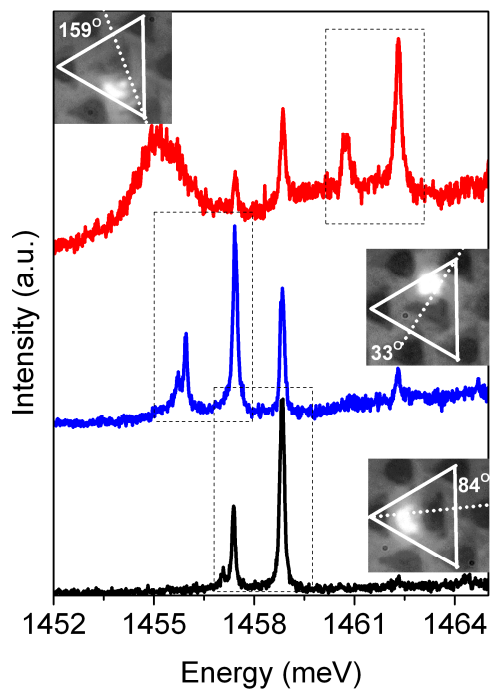

**Figure 7.** (Color online) Three spectra of three distinct CQDs located in the same pyramid. The spectra have been shifted vertically for clarity. For each spectrum, the position of the laser spot was optimized to maximize the PL signal from each CQDs. The bright spots in the inset optical micrographs correspond to the laser spot. The polarization axis of each CQD is shown in the insets.

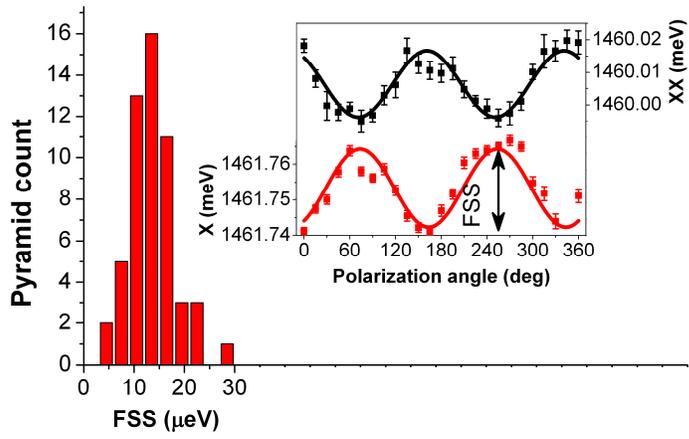

**Figure 8.** (Color online) The FSS distribution of CQDs. (Inset) Exciton (X) and biexciton (XX) of a CQD position as function of polarization angle.

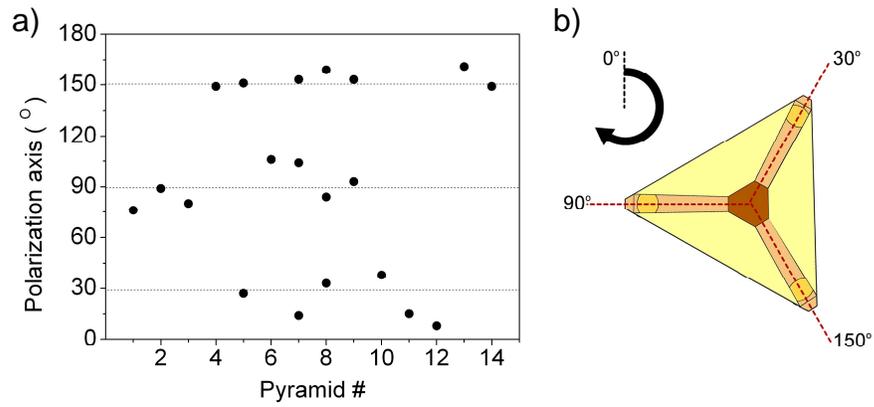

**Figure 9.** (Color online) (a) Distribution of CQDs polarization axis values. (b) Top-view projection of InGaAs heterostructures grown in a pyramidal recess. The angles represent orientation of LQWRs projections with respect to the zero axis in the experiment.

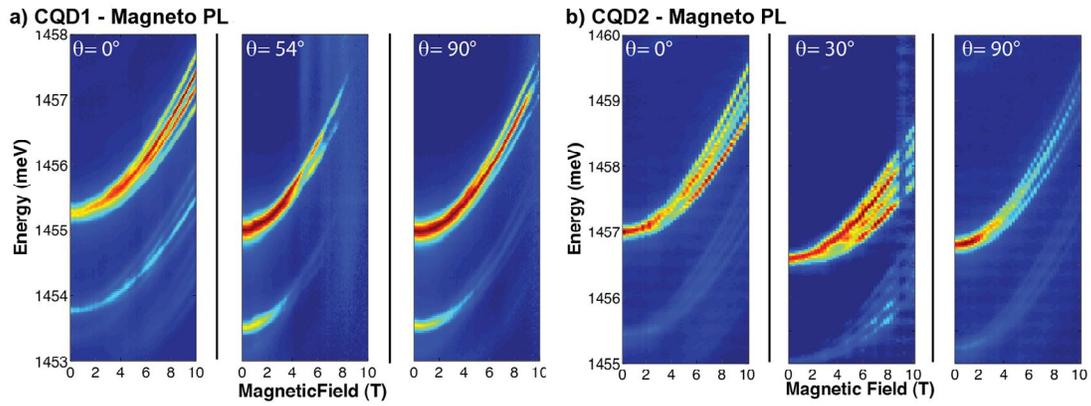

**Figure 10.** (Color online) (a) Magneto-PL scans for CQD1 taken for θ = 0, 54 and 90°. (b) Magneto-PL scans for CQD2 taken for θ = 0, 36° and 90°. In (a,b), the PL spectra are normalized and the intensity is colour-coded. Occasionally the diamagnetic forces in the apparatus alter the optical alignment of the system causing a loss of signal at fields larger than about 8 T.

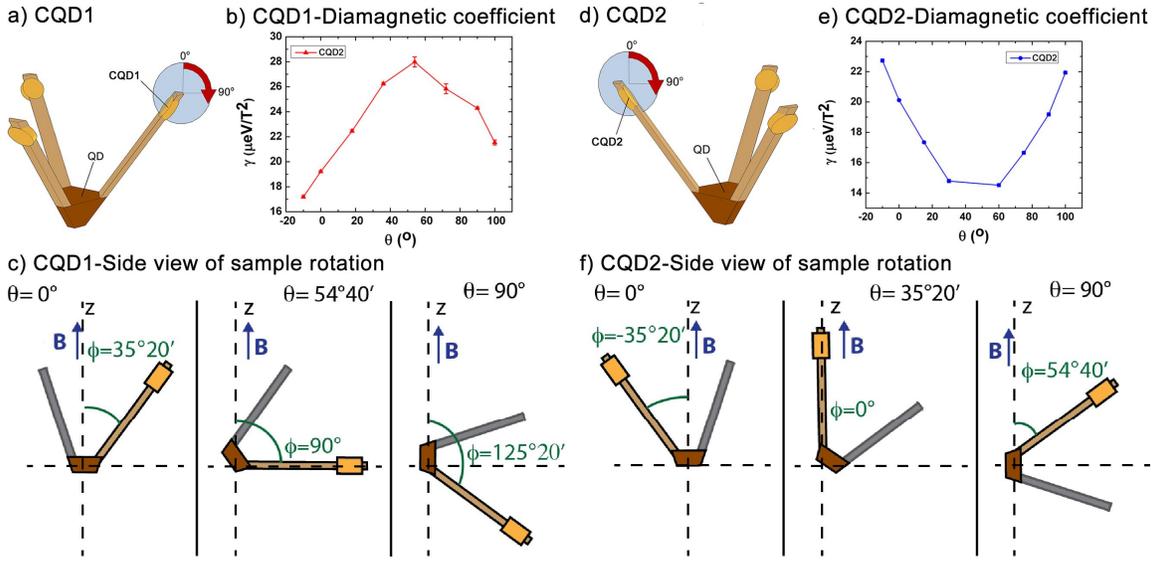

**Figure 11.** (Color online) (a) Sketch of CQD1 location within the ensemble of self-formed nanostructures. The sample is rotated clockwise (red arrow) and remains in the plane defined by the blue circle. (b) Evolution of the diamagnetic coefficient of CQD1 as a function of the angle θ between the sample and the magnetic field. The solid line is a guide to the eye. (c) Side-view of the lateral wires and CQD1 for θ = 0, 54°40' and 90°. The yellow wire contains CQD1 and remains in the rotation plane. The grey one represents the projection of the other two wires in the rotation plane. The angle φ shows the angle between the LQWR axis and the magnetic field. (d-f) Same as (a-c) for CQD2 and for θ = 0, 35°20' and 90°.

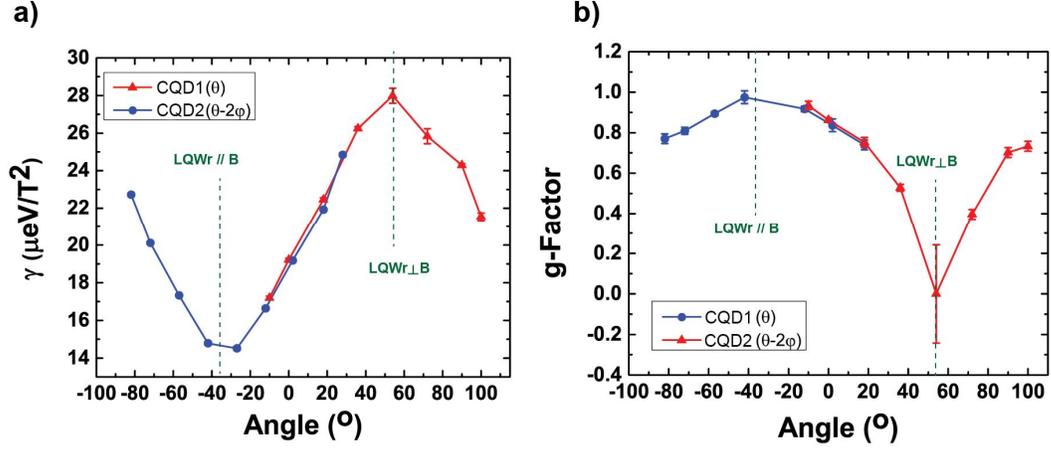

**Figure 12.** (Color online) (a) Exciton diamagnetic coefficient γ for CQD1 and CQD2 as a function of θ and (θ-2ϕ), respectively. (b) Bright exciton g-factor for CQD1 and CQD2 plotted as a function of θ and (θ-2ϕ), respectively. The large error bar for θ=54° arises from the fact that for this angle we could not resolve any splitting between the two exciton bright states. In both (a) and (b), solid lines are guides to the eyes.